\documentclass[twocolumn, prb, aps, floatfix]{revtex4}

\usepackage{graphicx}
\usepackage{color}

\begin{document}

\title{Efficient long-distance energy transport in molecular systems through adiabatic passage}

\author{Arend G. Dijkstra$^{a,b}$}
\author{Almut Beige$^b$}

\affiliation{$^{a}$The School of Chemistry, University of Leeds, Leeds LS2 9JT, United Kingdom}
\affiliation{$^{b}$The School of Physics and Astronomy, University of Leeds, Leeds LS2 9JT, United Kingdom}

\begin{abstract}
The efficiencies of light-harvesting complexes in biological systems can be much higher 
than the current efficiencies of artificial solar cells. In this paper, we therefore propose and analyse an energy transport mechanism which employs adiabatic passages between the states of an artificially designed antenna molecular 
system to significantly enhance the conversion of incoming light into internal energy. It is shown that the proposed transport mechanism is relatively robust against spontaneous emission and de-phasing, while also being able to take advantage of collective effects. Our aim is to provide new insight into the energy transport in molecular complexes and to improve the design of solar cells. 
\end{abstract}

\date{\today}

\maketitle

\section{Introduction}

Our ability to design molecular light harvesting systems is based on a detailed
understanding of light absorption, energy transport, charge separation and charge
transport processes. 
A key design criterion is the energy cost and size requirement of the molecular complexes.
In biological systems, charge separation in reaction centers is often optimized by
connecting these to antennae systems \cite{Blankenship.book} 
which maximize light absorption, but only
function in the presence of efficient energy transport pathways through the antennae
to the reaction centre. This process minimizes the number of reaction centres necessary.
In contrast to this, artificial light harvesting systems, like solar cells, follow different design principles.
While the charge separation and transport processes are also important for the overall
device efficiency of solar cells, the question remains whether functional antenna systems can be used to significantly improve their performance. Answering this question requires more insight into energy transport mechanisms in complex molecular systems.

A lot of work has gone into understanding energy transport pathways in molecular complexes.
On the larger scale, Foerster transfer theory and its multi-chromophoric 
version\cite{Jang.2004.prl.92.218301, Cleary.2013.pnas.110.8537} have
been used to explore energy transport pathways in systems such as light-harvesting
complex II from green plants.\cite{Bennett.2018.pnas.115.E9523}
Such work on energy transport in light-harvesting complexes often assumes that a single
quantum of excitation is present in the system (but see recent work in 
\onlinecite{Caycedo.2017.jpcl.8.6015}). 
A large amount of work has recently been done on the question whether such a quantum is
delocalised over more than one chromophore molecule. This question has been approached
with a combination of experimental tools using ultrafast laser pulses and theoretical
modelling. A lot of effort has focused on the dimer, but larger systems have been
studied as well, in particular, the Fenna-Matthews-Olson complex. 
It has been realised that non-Markovian effects are
important.\cite{Ishizaki.2009.jcp.130.234111} Subtle effects of the initial condition 
related to correlations
between electronic excitations and vibrations in nonlinear optical
experiments are usually ignored.\cite{Dijkstra.2012.njp.14.073027}

A key factor that is missing in this description of light-harvesting systems is the absorption process which
takes the system from the ground state of a donor molecule into the excited state of an acceptor molecule. Subtle quantum
effects are known to be able to populate certain states of the system selectively,
while leaving others empty.\cite{Beige1999,Beige2000,Facchi,Busch2010,Vitanov.2017.rmp.89.015006} A canonical example is
the $\Lambda $ system, in which excitation can be transferred directly via a Stimulated Raman Adiabatic Passage (STIRAP) from an initial 
to a final ground state.\cite{Eberly1984,Shore1998}  Even though these states are not directly coupled but only interact indirectly via a single excited state, this intermediate bridge does not acquire any population. A key advantage of such techniques is that they protect against loss mechanisms in the
intermediate state which is why they have found a wide range of applications, for example, in quantum technology.\cite{Kuhn2002,Kuhn2010,Gangloff.2019.science.364.62}

There are many ways of restricting the dynamics of open and closed quantum systems onto subspaces of states. What all of these mechanisms have in common is that they expose unwanted states to rapid dynamics. Rapid dynamics can be shown to effectively result in a quantum Zeno effect which suppresses the exchange of excitation between observed subspace.\cite{Facchi,Busch2010} For example, STIRAP works because the above described $\Lambda$ system possesses a zero energy eigenstate. The presence of strong interactions ensures that the $\Lambda$ system remains at all times in its zero eigenstate. When this state slowly changes in time, its system dynamics follows those changes adiabatically. Alternatively, continuous measurements can be used to restrict the dynamics of an open quantum system onto a higher-dimensional so-called decoherence-free subspace.\cite{Beige1999,Beige2000}

Using these ideas in the context of molecular complexes,
it should be possible to have direct transport of population from the ground state into an
acceptor molecule, which we define to be part of a product forming reaction centre,
under weak excitation using adiabatic passage. This means that
intermediate states are not significantly populated and that the mechanism is 
therefore not sensitive to dissipation. An important question for molecular systems
is how well the mechanism works in the presence of de-phasing, which is a key
ingredient of standard mechanisms of light harvesting complexes,\cite{Ishizaki.2009.pnas.106.17255} 
and is unavoidable in molecular systems at room temperature.

From work on energy transport, it is well known that 
factors that affect transport efficiency are the strength of the excitonic coupling
between molecules, as well as the fluctuations induced by the environment. 
Noise assisted transport has been introduced as a mechanism to explain how the interplay
between these two effects leads to transport that is more efficient than what is possible
in the purely coherent limit, where fluctuations are absent or in the overdamped limit,
where fluctuations are much stronger than excitonic coupling.\cite{Cao.2009.jpca.113.13825,
Rebentrost.2009.njp.11.033003,Caruso.2009.jcp.131.105106} By taking into account the ground state explicitly, 
we introduce another quantum effect that can be used to design efficient and resilient
energy transport networks in molecular materials.
We argue that this can lead to 
new mechanisms of energy transport that are not present when only excited states
are considered. We thereby introduce the process of energy transport through adiabatic passage
(ETAP). The mechanism of adiabatic passage doesn't work in a dimer, which shows that it is 
fundamentally different from noise assisted transport. 

Relevant to our model, 
in recent work it has been argued that at least three coupled molecules contribute
to the observed nonlinear optical signal in photosynthetic light harvesting 
complexes.\cite{Maiuri.2018.nchem.10.177} The same work also argues for the importance
of ground state vibrations for the interpretation of these experimental results. These
two arguments strengthen our motivation to explicitly include the ground state in
a model with three or more coupled chromophores. Overall, our approach could lead to new forms of coherent control of molecular 
excitations\cite{Tomasi.2018.arxiv.1810.03251}
as well as aiding the design of artificial light harvesting antennae
complexes with significantly increased efficiencies.

There are five sections in this paper. Section \ref{sec2} introduces the molecular structures and processes which we consider throughout this paper. Section \ref{sec3} discusses the main mechanism underlying energy transport through adiabatic passages. Section \ref{sec4} analysis the dynamics of the proposed molecular structures numerically. Finally, we review our findings in Section \ref{sec5}.

\section{Model} \label{sec2}

Next we introduce the relevant molecular systems and specify its Hamiltonians and noise models for the description of coherent and incoherent processes.

\subsection{A single-antenna molecular structure} 

\begin{figure}[t]
\includegraphics[width=8.5cm]{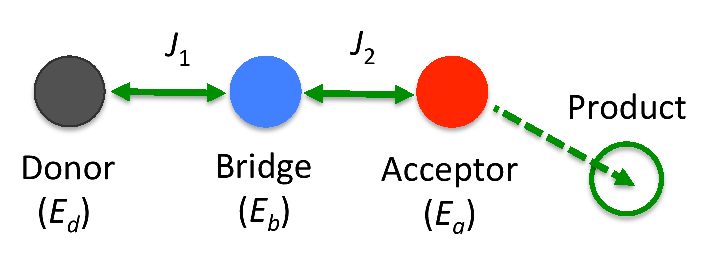} 
\caption{\label{fig:single} 
Cartoon of our single antenna model system. The donor molecule is excited
by light absorption from the ground state. It is coupled to a bridge, which is in turn
coupled to an acceptor. On the acceptor molecule, an incoherent process forms product. The
acceptor and product states together can be thought of as belonging to a reaction centre,
while the donor and bridge molecules form an antenna.}
\end{figure}

The simplest possible model that can be used to demonstrate the ETAP process is
a three-molecule system. Suppose there are three molecules, each of them containing a ground and an excited electronic state.  As usual, we ignore possible other electronic states as well as double excitation but we do explicitly include the ground state in our description. As illustrated in Fig.~\ref{fig:single}, our molecular system contains a donor molecule $(d)$, a bridge molecule $(b)$ and an acceptor molecule $(a)$.

In the following, we assume that the donor is excited from the ground
state $|g\rangle$ into an excited state $|d\rangle$ upon absorbing light, with an excitation energy $E_d$. During this excitation process, all other molecules remain in their respective ground states. For simplicity, we assume that the incoming light couples to the ground states of these molecules, although such couplings can easily be included in the model. We also assume that the goal of the energy transfer process is to reach a product state,
which is populated from the acceptor molecule with excited state $|a\rangle$ and excitation energy $E_a$ via an incoherent process. 
In-between the donor and the acceptor
is a bridge molecule (with excitation energy $E_b$ and excited state $|b\rangle$), 
which interacts with the donor and with the acceptor with coupling constants $J_1$ and $J_2$, respectively. Longer
chains with multiple bridge molecules are also possible, and have been investigated
in the context of electron transfer by other authors.\cite{Davis.1997.jpca.101.6158}
The purpose of the bridge is to allow excitation to travel over a longer distance. The dynamics of the excited quantum states of this model can be described by the Hamiltonian \cite{May.2008.jcp.129.114109}
\begin{equation} \label{H}
   H = \sum_{i=d,b,a} E_i \, |i \rangle \langle i| + J_1 \left( |b \rangle \langle d| + 
   {\rm H.c.} \right) + J_2 \left( |b \rangle \langle a| + {\rm H.c.} \right) .
\end{equation}
In addition to using this Hamiltonian, we generate a Liouville operator to propagate the density matrix $\rho$ of the above described molecular structure in the presence of de-phasing and spontaneous photon emission in the usual way.

More concretely, we assume that there is an incoherent process that takes population from the acceptor to generate a product. We also assume that this excitation is  trapped in this product state for long time compared to other processes. Excitation does not come back to the ground state within the time scale of our simulations. The incoherent product formation is modelled with a Lindblad term in the Liouville equation with  a pre-factor $\Gamma_P$. In addition, we describe de-phasing due to environment induced fluctuations 
by coupling the excited states of donor, acceptor, and bridge each to an independent
bath of harmonic oscillators, which modulates their respective excitation energy. The de-phasing
process outside the Markovian limit is modelled with a Drude-Lorentz spectral density, for
which the parameters are the re-organization energies $\Lambda $, the inverse bath time
scales $\gamma$ and the inverse temperature $\beta$. For simplicity, we assume that these
parameters are the same for each of the three baths, although this can easily be generalized. To be able to include de-phasing induced by a bath with arbitrary time scale, we use the hierarchical equations of motion (HEOM)\cite{Tanimura.2006.jpsj.75.082001} 
to propagate the system's dynamics. This formalism can handle
arbitrary time varying external fields which we explore to simulate different excitation conditions. In the following, we consider four different ways to excite the donor molecule.

{\bf Excitation model 1.} In excitation model 1, we consider continuous wave excitation with a laser with frequency
$\omega$ and field amplitude $A$. In this case, the interaction of the laser
with the system is modelled in the semi-classical approximation
by a time-dependent Hamiltonian $H'(t)$ of the form
\begin{equation} \label{hp1}
 H'(t) = -A \cos \omega t \, |d\rangle \langle g| + \textrm{H.c.}
\end{equation}
The amplitude $A$ should be understood as the inner product of the laser's electric field
amplitude $\vec E$
with the molecular transition dipole $\vec\mu$, $A = \vec\mu \cdot \vec E$. We will keep
this quantity constant, and we will also in most cases not vary $\omega$. Moreover H.c. denotes
the Hermitian conjugate. For weak excitation, the product population increases linearly with time after
an initial transient. In this linear regime, we can define a product formation rate 
\begin{equation} \label{R}
R = \mathrm{d}P(t)/\mathrm{d}t \, ,
\end{equation}
where $P(t)$ is the population on the product state. This rate is a natural measure for the efficiency of the system under continuous irradiation.

{\bf Excitation model 2.} In excitation model 2, we consider a laser pulse as is often used in ultrafast optical
experiments. The interaction Hamiltonian has the same form as in Eq.~(\ref{hp1}), but
the oscillating field now has an envelope, which we assume to be Gaussian with
center $t_0$ and standard deviation $\sigma$. In this case, the interaction Hamiltonian $H'(t)$ has
the form
\begin{equation} \label{hp2}
  H'(t) = -A \cos \omega t \, {\rm e}^{-(t-t_0)^2/2 \sigma^2} \, |d\rangle \langle g| + \textrm{H.c.} \, ,
\end{equation}
where we have absorbed the normalization factor of the Gaussian function into $A$. We
use a pulse half width of $\sigma=15$ fs. In this case, we quantify the performance of the system simply by measuring the amount
of product formed a long time after the pulse, which we take to be 26.5 ps. This measure
does not take into account the time it takes to form the product, but simply measures
the final amount.
To define a rate $R$ as well, we calculate 
\begin{equation} \label{R2}
R=1/\tau ~~ {\rm with} ~~ \tau=\int \mathrm{d}t \, (P_\mathrm{eq} - P(t)) \, .
\end{equation}
Here $\tau $ is the transport time and $P_\mathrm{eq}$ is the product population in equilibrium.

{\bf Excitation model 3.} In excitation model 3, we consider incoherent light, which is the natural excitation
condition in photosynthesis, as well as in most artificial light harvesting applications.
While we could in principle model this excitation process with the hierarchical equations
of motion, it is equally valid to use a rate of 
excitation $\Gamma$.\cite{Olsina.2014.arxiv.1408.5385} This
means that we add to the system's Liouville operator the (Lindblad) term
\begin{equation} \label{hp3}
  \mathcal{L}' \rho = \Gamma \, ( L \rho L^\dagger - \frac{1}{2} L^\dagger L \rho -
    \frac{1}{2} \rho L^\dagger L ), 
\end{equation} 
where $\rho$ is the density matrix and $L = |d\rangle \langle g|$. This
description of incoherent light does not take into account the superOhmic nature of the
spectral density expected for a photon bath.\cite{Pachon.2017.jpb.50.184003} In this case, we calculate the rate $R$ by taking the time derivative of the population in
the linear regime, as in excitation model 1. 

{\bf Excitation model 4.} Finally, excitation model 4, is introduced to benchmark all other excitation schemes and to show that it is indeed advantageous to include the donor ground state in energy transport simulations. As is often done in calculations, we consider the case where the ground
state is ignored, and the initial state of the system is simply chosen to be the state
with unit population in the donor excited state. If $\rho$ is the density matrix, this
means that $\langle d| \rho | d \rangle = 1$ initially, while all other matrix elements of the density
matrix are zero. The Franck Condon principle is also applied, so that the initial excitonic
density matrix is not correlated with the vibrational bath. In this case, we calculate the rate $R$ as in excitation model 2. 

\begin{figure*}[t]
\includegraphics[width=17cm]{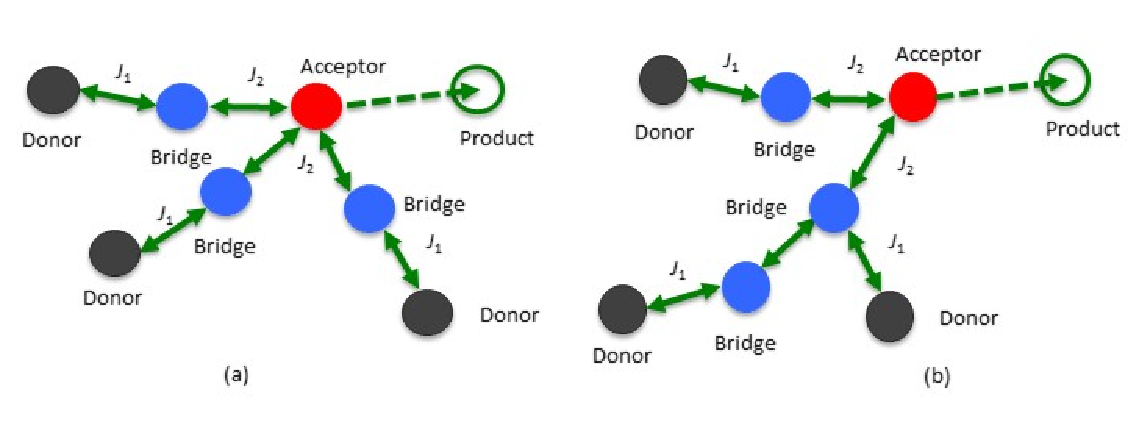} 
\caption{\label{fig:multiple} 
Systems with multiple antennae can be constructed as well. This figure shows two examples.
(a) multiple bridges and (b) including a longer distance bridge.}
\end{figure*}

\subsection{Multi-antenna molecular structures} 

A straightforward variation of the molecular structure which we discussed in the previous 
subsection is the addition of more antennae, as illustrated in Fig.~\ref{fig:multiple}. 
This allows the reaction centre to receive more excitation as long as the incoming photon
flux is relatively weak, which improves the system if reaction centres are more complicated to construct than antennae. Nature exploits this
strategy, for example in light harvesting in green plants.

In our multi-antenna complex, there are multiple donors, which are each coupled to their
own independent bridges. These bridges are then all coupled to a single acceptor, where
the product formation takes place incoherently. The logic behind having a single acceptor
molecule rather than independent acceptors for each antenna is that the acceptor is considered
to be part of the reaction centre. For simplicity, we take the excitation energies of all
donors to be equal, and assume equal site energies 
for all bridge molecules. Couplings $J_1$ and $J_2$ and bath
parameters are also chosen to be replicated in each antenna.

\section{Energy transfer through adiabatic passage (ETAP)} \label{sec3}

Our system is inspired by multi-state Stimulated Raman Adiabatic Passage (STIRAP) chains,\cite{Busch2010,Vitanov.2017.rmp.89.015006} of which three-level $\Lambda $ systems are the simplest example.\cite{Eberly1984,Shore1998} One application of STIRAP is to generate single photons on demand by mapping the ground state of an atom inside an optical cavity onto excitation in the free radiation field.\cite{Kuhn2002,Kuhn2010} Here we use adiabatic passages to realise the inverse process. Our aim is not to generate light but to guide incoming light with a very high efficiency to the ground state of a molecular reaction centre.

\subsection{A single-antenna molecular structure}

First we have a closer look at the energy transport within the single-antenna molecular structure shown in Fig.~\ref{fig:single}. The first condition we need for the ETAP process to work is
\begin{equation} \label{condi}
A \ll J_1, \,  J_2 \, , 
\end{equation}
where $A$, $J_1$ and $J_2$ are the coupling constants which we introduced in Eqs.~(\ref{H})--(\ref{hp2}). 
Let us assume for a moment that $A = 0$ and $E_a - \omega = E_d - \omega = 0$. In this case, the Hamiltonian $H$ in Eq.~(\ref{H}) possesses two zero energy eigenstates, $|\lambda_1 \rangle$ and $|\lambda_2 \rangle$, which are given by
\begin{equation} \label{vectors}
|\lambda_1 \rangle = |g \rangle ~~ {\rm and} ~~
|\lambda_2 \rangle = {1 \over \sqrt{J_1^2 + J_2^2}} \left( J_1 \, |a \rangle - J_2 \, |d \rangle \right) \, .
\end{equation} 
All other energy eigenstates of $H$ evolve relatively rapidly in time. Taking this into account, when studying the effect of a relatively weak interaction $H'(t)$ and adiabatically eliminating all states which evolve rapidly in time from the system dynamics, one can show that our molecular structure evolves to a very good approximation according to the effective Hamiltonian\cite{Busch2010} 
\begin{equation} \label{effective}
H_{\rm eff}(t) = I \!\! P \, H'(t) \, I \!\! P 
\end{equation}
with $I \!\! P = |\lambda_1 \rangle \langle \lambda_1| + |\lambda_2 \rangle \langle \lambda_2|$. The dynamics of the system remains restricted onto a decoherence-free subspace of slowly evolving states.\cite{Beige1999,Beige2000} For example, in case of excitation model 1, the interaction in Eq.~(\ref{hp1}) results in the effective Hamiltonian 
\begin{equation} \label{effective2}
H_{\rm eff}(t) = A_{\rm eff} \, \cos \omega t \, |\lambda_2 \rangle \langle g| + \textrm{H.c.} ~~ {\rm with} ~~ A_{\rm eff} = {A J_2 \over \sqrt{J_1^2 + J_2^2}} \, .
\end{equation}
The applied laser field couples the ground state $|g \rangle$ of the donor directly to the excited states of donor and acceptor. If we choose 
\begin{equation} \label{condi2}
J_2 \ll  J_1 \, , 
\end{equation}
we can achieve an almost direct coupling between the ground state of the donor and the excited state of the acceptor molecule. This coupling comes at the expense of a strongly reduced effective coupling rate $A_{\rm eff}$ but also minimises the population of intermediate excited states. Although it might take longer for the incoming light to arrive at the product, the overall energy transfer can become highly efficient with almost all available excitation being transported to the center. Moreover, the energy transfer becomes highly insensitive of de-phasing and other forms of decoherence. In fact, adiabatic processes are not only used to generate single photons on demand, they are also used to aid the generation of entangled states and to protect quantum computing against dissipation.\cite{Beige1999,Beige2000} As we shall see in the next section, optimising ETAP in light harvesting complexes requires a careful optimisation of all system parameters, including detunings, but can indeed result in a significant increase of overall transfer rates.

\subsection{Multi-antenna molecular structures}

For the multi-antenna molecular complex, one could naively expect that the product formation rate
will scale simply linearly with the number of antennae. While this is often a good
first-order approximation, it ignores the effects of de-localization across the different
antennae, which is relevant here because the whole complex is still smaller than the wavelength
of light and can therefore be coherently excited. For simplicity, we assume in the following that all antennae experience the same excitation process
and the same coupling constants. Unevenly distributed excitation of donor molecules might reduce the efficiency of the proposed energy transfer.\cite{Cao2013}

Suppose a light-harvesting molecular structure contains $N$ acceptor molecules which all link via a single bridge to a donor as shown in Fig.~\ref{fig:multiple}(a). In this case, the Hamiltonian $H$ in Eq.~(\ref{H}) changes into 
\begin{eqnarray} \label{Hrep}
 H &=& \sum_{n=1}^N \sum_{i=d,b,a} E_i \, |i_n \rangle \langle i_n| + J_1 \left( |b_n \rangle \langle d_{n}| + {\rm H.c.} \right) \nonumber \\
   &+& J_2 \left( |b_n \rangle \langle a| + {\rm H.c.} \right) ,
\end{eqnarray}
where $n$ indicates which antenna a certain state belongs to. Moreover, in case of excitation model 1, the laser interaction Hamiltonian $H'(t)$ in Eq.~(\ref{hp1}) becomes
\begin{equation} \label{hp1many}
 H'(t) = - \sum_{n=1}^N A \cos \omega t \, |d_n \rangle \langle g| + \textrm{H.c.} \, ,
\end{equation}
where $|g \rangle$ denotes the state with all donor molecules in their respective ground state. However, in the weak excitation limit, both Hamiltonians can be shown to reduce effectively to the Hamiltonians in Eqs.~(\ref{H}) and (\ref{hp1}). All we need to do is to replace the single antenna states $|d \rangle$ and $|b \rangle$ by the (normalised) Dicke states\cite{Dicke1954}
\begin{equation} \label{Dicke}
|D \rangle = {1 \over \sqrt{N}} \sum_{n=1}^N |d_n \rangle  ~~ {\rm and} ~~
|B \rangle = {1 \over \sqrt{N}} \sum_{n=1}^N |b_n \rangle 
\end{equation}
and the coupling constants $A$ and $J_2$ by $\sqrt{N} A$ and $\sqrt{N} J_2$, respectively. Hence evenly driven molecular structures with multiple antennas experience the same dynamics as the single-antenna structure in Fig.~\ref{fig:single}, while coupling constants are collectively enhanced. This needs to be taken into account when optimising energy transfer processes in light-harvesting systems and can result in a further increase of efficiency.

\begin{figure}[t]
\includegraphics{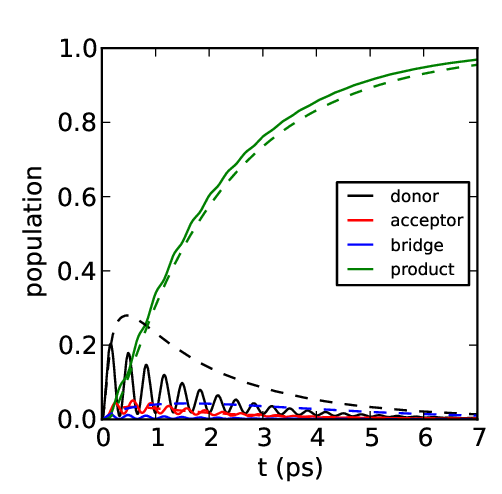} 
\caption{\label{fig:etapdemo}
State populations as a function of waiting time for continuous wave excitation of
a single antenna model system. 
The excitation frequency was set to the lowest eigenvalue
of the system Hamiltonian, which is 17.988 kcm$^{-1}$. Dashed lines include de-phasing
on all three molecules while for the solid lines $\Lambda =0$. Population on the 
bridge molecule is small, so losses there will not affect the process much.}
\end{figure}

\section{Results} \label{sec4}

The standard single antenna system that we investigate has a donor which is 
slightly blue-shifted with respect to
the acceptor, providing downhill energy transport. We include a bridge with an energy above
the donor and acceptor energies, so that the three excited states form a $\Lambda $ system.
Later, we will investigate the effect of the bridge energy on the product formation.
For now, the parameters are $E_d = 18.2$ kcm$^{-1}$,
$E_a = 18.0$ kcm$^{-1}$, $E_b=19.0$ kcm$^{-1}$, $J_1 = 0.5$ kcm$^{-1}$ and 
$J_2 = 0.1$ kcm$^{-1}$. The product
is formed from the acceptor in an incoherent process with $\Gamma_P = 0.25$ kcm$^{-1}$. 
The de-phasing parameters are $\Lambda = 0.05$ kcm$^{-1}$, $\gamma = 0.15$ kcm$^{-1}$ and
$\beta = 5.0 \cdot 10^{-3}$ cm. Although these parameters are not meant to model a concrete system,
they are all chosen to be realistic for molecular complexes. 
Because $k_B T = 1/\beta$ is on the order of or smaller than
the energy gaps in the system and $\beta \gamma < 1$, we do not include Matsubara frequency
terms in our HEOM propagation.

The first result is a demonstration of the ETAP process with excitation model 1. 
In Figure~\ref{fig:etapdemo}, we observe that the bridge state never has large
population. This is the essential feature of the process that makes the energy transfer
over longer distances possible with only small losses.

We will now turn our attention to excitation with a short pulse (excitation model 2). 
The base line scenario that we compare our antenna system with is a single reaction centre,
that is, a molecule that absorbs light and generates product.

First, we compare this reaction centre with a complex with a single antenna, as is shown in
figure~\ref{fig:single}. 
With a pulse of 15 fs halve width 
(standard deviation of Gaussian pulse) and 0.05 kcm$^{-1}$ amplitude, the total amount
of product produced after a long time (26.5 ps) is calculated for this system by integrating
the equations of motion numerically.
In these calculations, the bridge energies were set to 19.0 kcm$^{-1}$ and the excitation
centre energy was $\omega = $ 18.0 kcm$^{-1}$. For the single
reaction centre, the total product is 0.021. Remarkably, adding an antenna slightly
\emph{increases} the total product to 0.022, even though only the donor molecule is 
optically excited. We have confirmed that the amount of product formed hardly depends
on the time scale of the environment (the value of $\gamma$). We find that the amount
of product produced does depend on the strength of de-phasing ($\Lambda $), but that even
for large values of $\Lambda $ product is still formed. In numbers, the amount of
product varies between 0.023 for $\Lambda =25 \mathrm{cm}^{-1}$ and 0.012 for the large
value of $\Lambda =500 \mathrm{cm}^{-1}$ (the latter value is on the same order as the largest
couplings). 
We can therefore say that the product formation
process is robust against de-phasing in a parameter regime relevant for realistic systems, as
has been suggested before in the context of STIRAP,\cite{Shi.2003.jcp.119.11773}
even though it is known in general that adiabatic passage can be sensitive to de-phasing.
\cite{Vitanov.2017.rmp.89.015006}
The time it takes to form the product after
the pulse also depends on the de-phasing strength. 

For a setup with three antennae (Figure~\ref{fig:multiple} (a)), and otherwise the same parameters,
we find a product formation of 0.065,
almost as good as three times the production with one arm. This shows that the mechanism
proposed here also works in extended systems. These have the advantage that only
a single reaction centre is needed that can function with multiple antennae. Our calculation
results shows that this setup will function efficiently with three antennae. It does
not exclude the possibility that other setups with more antennae may work as well. In
practice, geometry constraints may apply, and longer bridges (Figure~\ref{fig:multiple} (b)) 
could be investigated in larger systems.

\begin{figure}[t]
\includegraphics{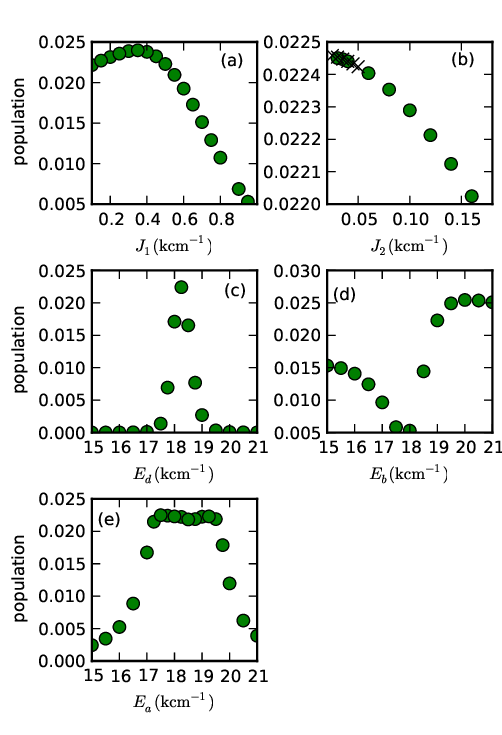} 
\caption{\label{fig:paramfinal}
Product formation in the model system with a single antenna as a function
of the Hamiltonian parameters. In panel (b), crosses indicate simulation results for
a twice longer time after the pulse, 53 ps, while 26.5 ps was used for all simulation
results shown as filled circles.}
\end{figure}

Next, we go back to the single antenna and 
investigate the Hamiltonian parameters that optimize the performance of our 
model system. In figure \ref{fig:paramfinal}, we plot the amount of product formed after
excitation with a short pulse, as a function of the couplings and site energies that
appear in the Hamiltonian. We scan one parameter at a time, while keeping the others
fixed at the values mentioned above. For the coupling $J_1$ (panel (a)), 
there is an optimal value, while
the performance of the system keeps increasing as $J_2$ is decreased (panel (b)). 
In this regime, however,
it takes a very long time to reach the product. As a function of the donor energy $E_d$,
there is a narrow peak in performance (panel (c)), 
which reflects the effective absorption of light. 

The
population produced as a function of bridge energy shows an interesting behaviour. We find
that the presence of a bridge with an energy offset strongly increases the efficiency of 
the system (panel (d)). The mechanism
also functions if the bridge energy is below the donor and acceptor energies (a valley), although
not as well as with a positive bridge energy, but
\emph{far less} product is formed if the bridge is low, i.e. if the bridge energy is
close to the donor and acceptor energies. Choosing the optimal bridge energy enhances
the product formation by a factor of almost five. We have also confirmed (data not shown)
that the ratio of the time-integrated bridge population and the time-integrated donor
population shows a minimum close to the bridge energy that leads to optimal
product formation. The absolute value of the integrated bridge population also has a 
minimum in this
region. These observations confirm that using a bridge with an energy offset does not
only optimize the product formation, but also leads to losses from the intermediate bridge
state that are as small as possible. They are in line with the signatures of 
STIRAP.\cite{Vitanov.2017.rmp.89.015006} Finally, we find that our model system is
not very sensitive to the acceptor energy, but that a broad plateau of values exists for
which product formation is efficient (panel (e)).

\begin{figure}[t]
\includegraphics{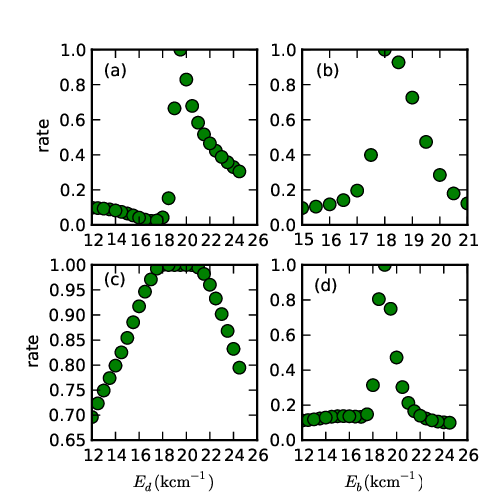} 
\caption{\label{fig:rate}
Normalized rates as a function of bridge energy for the four different excitation
conditions. Panel (a): excitation condition 1 (continuous wave irradiation), panel (b):
excitation condition 2 (pulsed excitation), panel (c): excitation condition 3 (incoherent
light), panel(d): excitation condition 4 (initial population on donor excited state). Rates
were calculated as described in the Model section.}
\end{figure}

Because the most salient feature of the results presented in the previous paragraph is
the necessity of a bridge with an energy offset 
to optimize performance, we now investigate how the bridge energy
affects transport rates for different excitation conditions.
In 
Figure~\ref{fig:rate} we plot the rate
of product formation versus the bridge energy. From panel (b), we see that the rate does
not capture the large loss in product formation when there is no bridge present. Instead,
without a bridge the population transport is fastest. This shows that just using the rate
of product formation in this model with pulsed excitation does not necessarily reveal
the optimal system parameters for light harvesting.

We also observe that for incoherent light excitation (panel (c)), the rate of product formation does
not sensitively depend on the bridge energy, but shows a plateau. For coherent continuous
wave excitation (panel (a)), however, there is a sharp maximum in the rate. We also find that for the
calculation of the rate, excitation condition 4 (panel (d)) 
with an initial population of one on the
donor gives different results from excitation with incoherent light (excitation condition 3).

\section{Conclusions} \label{sec5}

In this work, we have discussed model systems for light harvesting inspired by adiabatic
passage. We have found that for pulsed excitation, the presence of a bridge in the system
strongly enhances the amount of product formed in the system. Systems with multiple
antennae also function efficiently. The product formation rate
under excitation with incoherent light does not strongly depend on the bridge energy. The
model systems that we have studied allow for efficient long-range energy transport in
molecular complexes that is relatively insensitive to decay and de-phasing processes.

A key ingredient that we have not included in our present model are vibronic effects, that
is, underdamped vibrations displaced in the excited electronic state. Such vibrations
can help energy transport through resonances\cite{Dijkstra.2015.jpcl.6.627} 
and are generally important
in understanding the energy transport and spectroscopy of molecular 
systems.\cite{Thyrhaug.2018.nchem.10.780} While
vibronic effects are beyond the scope of the current work, they can easily be included
by using hierarchical equations of motion for the underdamped Brownian oscillator spectral
density.\cite{Tanaka.2009.jpsj.78.073802} Other forms of the spectral density for
incoherent light have also been suggested,\cite{Pachon.2017.jpb.50.184003} but a detailed 
investigation of such effects is beyond the scope of this work. It would also be interesting
to investigate how our model system can be studied experimentally with ultrafast
nonlinear optical spectroscopy\cite{Do.2017.jcp.147.144103} or time-resolved 
fluorescence.\cite{Tempelaar.2014.jpcl.5.1505} We expect that our work will stimulate the investigation of efficient man-made antennae
complexes in materials such as fluorographene\cite{Slama.2018.arxiv.1801.08509} or
dendrimer molecules.\cite{Supritz.2006.jlumin.119.337} \\[0.5cm]
{\bf Acknowledgement.} AB acknowledges financial support from the Oxford Quantum Technology Hub NQIT (grant number EP/M013243/1). Statement of compliance with EPSRC policy framework on research data: This publication is theoretical work that does not require supporting research data.

\end{document}